\documentclass[final,5p,times,twocolumn]{elsarticle}

\usepackage{amsmath}
\usepackage{amssymb}
\usepackage{graphicx}
\usepackage{color}

\usepackage[mathscr,scaled=1.15]{urwchancal}
\DeclareFontFamily{OT1}{pzc}{}
\DeclareFontShape{OT1}{pzc}{m}{it}%
{<-> s * [1.15] pzcmi7t}{}
\DeclareMathAlphabet{\mathpzc}{OT1}{pzc}{m}{it}

\biboptions{sort&compress}

\journal{Physics Letters B}

\begin{document}

\begin{frontmatter}

\title{Understanding the nucleon as a Borromean bound-state}

\author[USAL]{Jorge Segovia}
\ead{segonza@usal.es}
\author[ANL]{Craig D. Roberts}
\ead{cdroberts@anl.gov}
\author[JARA]{Sebastian M. Schmidt}
\ead{s.schmidt@fz-juelich.de}
\address[USAL]{Instituto Universitario de F\'isica Fundamental y Matem\'aticas
(IUFFyM), Universidad de Salamanca, E-37008 Salamanca, Spain}

\address[ANL]{Physics Division, Argonne National Laboratory, Argonne, Illinois
60439, USA}

\address[JARA]{Institute for Advanced Simulation, Forschungszentrum J\"ulich and JARA, D-52425 J\"ulich, Germany}
%
\date{11 June 2015}

\begin{abstract}
%
Analyses of the three valence-quark bound-state problem in relativistic quantum field theory predict that the nucleon may be understood primarily as a Borromean bound-state, in which binding arises mainly from two separate effects.  One originates in non-Abelian facets of QCD that are expressed in the strong running coupling and generate confined but strongly-correlated colour-antitriplet diquark clusters in both the scalar-isoscalar and pseudovector-isotriplet channels.  That attraction is magnified by quark exchange associated with diquark breakup and reformation.  Diquark clustering is driven by the same mechanism which dynamically breaks chiral symmetry in the Standard Model.  It has numerous observable consequences, the complete elucidation of which requires a framework that also simultaneously expresses the running of the coupling and masses in the strong interaction.  Planned experiments are capable of validating this picture.
\end{abstract}


\end{frontmatter}


\noindent\textbf{1.$\;$Introduction}.
The proton is the core of the hydrogen atom, lies at the heart of every nucleus, and has never been observed to decay; but it is nevertheless a composite object, whose properties and interactions are determined by its valence-quark content: $u$ + $u$ + $d$, \emph{i.e}.\ two up ($u$) quarks and one down ($d$) quark.  So far as is now known \cite{Agashe:2014kda}, bound-states seeded by two valence-quarks do not exist; and the only two-body composites are those associated with a valence-quark and -antiquark, \emph{i.e}.\ mesons.  These features are supposed to derive from colour confinement. Suspected to emerge in QCD, confinement is an empirical reality; but there is no universally agreed theoretical understanding.

Such observations lead one to a position from which the proton may be viewed as a Borromean bound-state, \emph{viz}.\ a system constituted from three bodies, no two of which can combine to produce an independent, asymptotic two-body bound-state.  In QCD the complete picture of the proton is more complicated, owing, in large part, to the loss of particle number conservation in quantum field theory and the concomitant frame- and scale-dependence of any Fock space expansion of the proton's wave function \cite{Dirac:1949cp, Keister:1991sb, Coester:1992cg, Brodsky:1997de}.  Notwithstanding that, the Borromean analogy provides an instructive perspective from which to consider both quantum mechanical models and continuum treatments of the nucleon bound-state problem in QCD.  It poses a crucial question: Whence binding between the valence quarks in the proton, \emph{i.e}.\ what holds the proton together?

In numerical simulations of lattice-regularised QCD (lQCD) that use static sources to represent the proton's valence-quarks, a ``Y-junction'' flux-tube picture of nucleon structure is produced, \emph{e.g}.\ Ref.\,\cite{Bissey:2006bz, Bissey:2009gw}.  This might be viewed as originating in the three-gluon vertex, which signals the non-Abelian character of QCD and is the source of asymptotic freedom \cite{Politzer:2005kc, Gross:2005kv, Wilczek:2005az}.  Such results and notions would suggest a key role for the three-gluon vertex in nucleon structure \emph{if} they were equally valid in real-world QCD wherein light dynamical quarks are ubiquitous.  As will become evident, however, they are not; and so a different explanation of binding within the nucleon must be found.

\smallskip

\noindent\textbf{2.$\;$DCSB and diquark correlations}.
Dynamical chiral symmetry breaking (DCSB) is another of QCD's emergent phenomena; and contemporary theory indicates that it is responsible for more than 98\% of the visible mass in the Universe \cite{national2012NuclearS, Brodsky:2015aia}.  We judge it probable that DCSB and confinement, defined via the violation of reflection positivity by coloured Schwinger functions (see, \emph{e.g}.\ Refs.\,\cite{Gribov:1999ui, Munczek:1983dx, Stingl:1983pt, Cahill:1988zi, Krein:1990sf} and citations thereof) have a common origin in the Standard Model; but this does not mean that DCSB and confinement must necessarily appear together.  Models can readily be built that express one without the other, \emph{e.g}.\ numerous constituent quark models express confinement through potentials that rise rapidly with interparticle separation but nevertheless possess no ready definition of a chiral limit; and models of the Nambu--Jona-Lasinio type typically express DCSB but not confinement.

DCSB ensures the existence of nearly-massless pseudo-Goldstone modes (pions), each constituted from a valence-quark and -antiquark whose individual Lagrangian current-quark masses are $<1$\% of the proton mass \cite{Maris:1997hd}.  In the presence of these modes, no flux tube between a static colour source and sink can have a measurable existence.  To verify this statement, consider such a tube being stretched between a source and sink.  The potential energy accumulated within the tube may increase only until it reaches that required to produce a particle-antiparticle pair of the theory's pseudo-Goldstone modes.  Simulations of lQCD show \cite{Bali:2005fuS, Prkacin:2005dc} that the flux tube then disappears instantaneously along its entire length, leaving two isolated colour-singlet systems.  The length-scale associated with this effect in QCD is $r_{\not\sigma} \simeq (1/3)\,$fm and hence if any such string forms, it would dissolve well within hadron interiors.

This discussion has exposed two corollaries of DCSB that are crucial in determining the observable features of the Standard Model.  Another equally important consequence of DCSB is less well known.  Namely, any interaction capable of creating pseudo-Goldstone modes as bound-states of a light dressed-quark and -antiquark, and reproducing the measured value of their leptonic decay constants, will necessarily also generate strong colour-antitriplet correlations between any two dressed quarks contained within a nucleon.  Although a rigorous proof within QCD cannot be claimed, this assertion is based upon an accumulated body of evidence, gathered in two decades of studying two- and three-body bound-state problems in hadron physics, \emph{e.g}.\ Refs.\,\cite{Cahill:1987qr, Cahill:1988dx, Bender:1996bb, Oettel:1998bk, Bloch:1999vk, Maris:2002yu, Bender:2002as, Bhagwat:2004hn, Cloet:2008re, Eichmann:2008ef, Eichmann:2009qa, Roberts:2011cf, Eichmann:2011aa, Segovia:2014aza, Segovia:2015hra}.  No realistic counter examples are known; and the existence of such diquark correlations is also supported by simulations of lQCD \cite{Alexandrou:2006cq, Babich:2007ah}.

The properties of diquark correlations have been charted.  Most importantly, diquarks are confined.  However, this is not true if the leading-order (rainbow-ladder, RL \cite{Munczek:1994zz, Bender:1996bb}) truncation is used to define the associated scattering problem \cite{Maris:2002yu}.  Corrections to that simplest symmetry-preserving approximation are critical in quark-quark channels: they eliminate bound-state poles from the quark-quark scattering matrix but preserve the strong correlations \cite{Bender:1996bb, Bender:2002as, Bhagwat:2004hn}.

Additionally, owing to properties of charge-conjugation, a diquark with spin-parity $J^P$ may be viewed as a partner to the analogous $J^{-P}$ meson \cite{Cahill:1987qr}.  It follows that scalar, isospin-zero and pseudovector, isospin-one diquark correlations are the strongest; and whilst no pole-mass exists, the following mass-scales, which express the strength and range of the correlation and are each bounded below by the partnered meson's mass, may be associated with these diquarks \cite{Cahill:1987qr, Maris:2002yu, Alexandrou:2006cq, Babich:2007ah}: $m_{[ud]_{0^+}} \approx 0.7-0.8\,$GeV, $m_{\{uu\}_{1^+}}  \approx 0.9-1.1\,$GeV,
with $m_{\{dd\}_{1^+}}=m_{\{ud\}_{1^+}} = m_{\{uu\}_{1^+}}$ in the isospin symmetric limit.
Realistic diquark correlations are also soft.  They possess an electromagnetic size that is bounded below by that of the analogous mesonic system, \emph{viz}.\ \cite{Maris:2004bp, Roberts:2011wyS}:
\begin{equation}
\label{qqradii}
r_{[ud]_{0^+}} \gtrsim r_\pi\,, \quad
r_{\{uu\}_{1^+}} \gtrsim r_\rho\,,
\end{equation}
with $r_{\{uu\}_{1^+}} > r_{[ud]_{0^+}}$.  As in the meson sector, these scales are all set by that associated with DCSB.

It is worth remarking here that in a dynamical theory based on SU$(2)$-colour, diquarks are colour-singlets.  They would thus exist as asymptotic states and form mass-degenerate multiplets with mesons composed from like-flavoured quarks.  (These properties are a manifestation of Pauli-G\"ursey symmetry \cite{Pauli:1957, Gursey:1958}.)  Consequently, the $[ud]_{0^+}$ diquark would be massless in the presence of DCSB, matching the pion, and the $\{ud\}_{1^+}$ diquark would be degenerate with the theory's $\rho$-meson.  Such identities are lost in changing the gauge group to SU$(3)$-colour; but clear and instructive similarities between mesons and diquarks nevertheless remain, as we have described above.

\smallskip

\noindent\textbf{3.$\;$Diquarks in the nucleon}.
%
The bulk of QCD's particular features and nonperturbative phenomena can be traced to the evolution of the strong running coupling.  Its unique characteristics are primarily determined by the three-gluon vertex: the four-gluon vertex does not contribute dynamically at leading order in perturbative analyses of matrix elements; and nonperturbative continuum analyses of QCD's gauge sector indicate that satisfactory agreement with gluon propagator results from lQCD simulations is typically obtained without reference to dynamical contributions from the four-gluon vertex, \emph{e.g}.\ Refs.\,\cite{Aguilar:2008xm, Aguilar:2009ke, Aguilar:2009nf, Maas:2011se, Boucaud:2011ug, Pennington:2011xs, Binosi:2012sj, Strauss:2012dg, Meyers:2014iwa}.  The three-gluon vertex is therefore the dominant factor in producing the class of renormalisation-group-invariant running interactions that have provided both successful descriptions of and predictions for many hadron observables \cite{Maris:2003vk, Chang:2011vu, Bashir:2012fs, Cloet:2013jya, Binosi:2014aea}.  It is this class of interactions that generates the strong attraction between two quarks which produces tight diquark correlations in analyses of the three valence-quark scattering problem.

\begin{figure}[!t]
\centerline{%
\includegraphics[clip,width=0.4\textwidth]{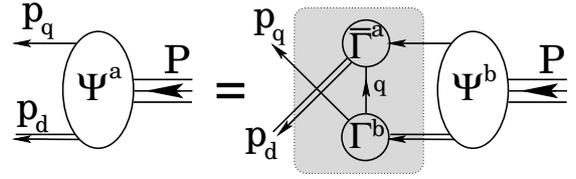}}
\caption{\label{fig:Faddeev} Poincar\'e covariant Faddeev equation.  $\Psi$ is the Faddeev amplitude for a baryon of total momentum $P= p_q + p_d$, where $p_{q,d}$ are, respectively, the momenta of the quark and diquark within the bound-state.  The shaded area demarcates the Faddeev equation kernel: \emph{single line},
dressed-quark propagator; $\Gamma$,  diquark correlation amplitude; and \emph{double line}, diquark propagator.}
\end{figure}

The existence of tight diquark correlations considerably simplifies analyses of the three valence-quark scattering problem and hence baryon bound states because it reduces that task to solving a Poincar\'e covariant Faddeev equation \cite{Cahill:1988dx}, depicted in Fig.\,\ref{fig:Faddeev}.  The three gluon vertex is not explicitly part of the bound-state kernel in this picture of the nucleon.  Instead, one capitalises on the fact that phase-space factors materially enhance two-body interactions over $n\geq 3$-body interactions and exploits the dominant role played by diquark correlations in the two-body subsystems.  Then, whilst an explicit three-body term might affect fine details of baryon structure, the dominant effect of non-Abelian multi-gluon vertices is expressed in the formation of diquark correlations.  Such a nucleon is then a compound system whose properties and interactions are primarily determined by the quark$+$diquark structure evident in Fig.\,\ref{fig:Faddeev}.

It is important to highlight that both scalar-isoscalar and pseudovector-isotriplet diquark correlations feature within a nucleon. Any study that neglects pseudovector diquarks is unrealistic because no self-consistent solution of the Faddeev equation in Fig.\,\ref{fig:Faddeev} can produce a nucleon constructed solely from a scalar diquark, \emph{e.g}.\ pseudovector diquarks typically provide roughly 150\,MeV of attraction \cite{Roberts:2011cf}.  The relative probability of scalar versus pseudovector diquarks in a nucleon is a dynamical statement.  Realistic computations predict a scalar diquark strength of approximately 60\% \cite{Cloet:2008re, Segovia:2014aza, Segovia:2015hra}.  As will become clear, this prediction can be tested by contemporary experiments.

\begin{figure}[!t]
\centerline{%
\includegraphics[width=0.4\textwidth]{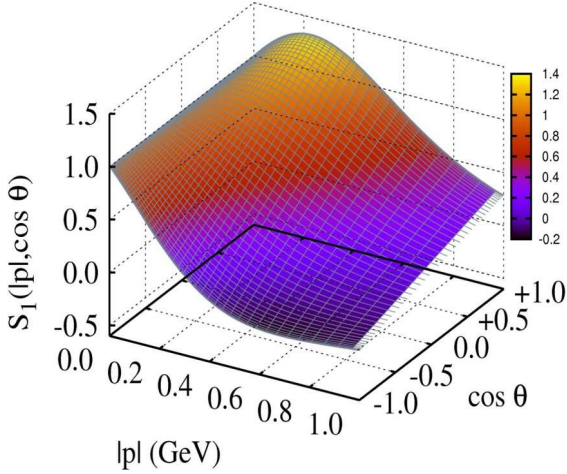}}
\caption{\label{figS1} Dominant piece in the nucleon's eight-component Poincar\'e-covariant Faddeev amplitude: $s_1(|p|,\cos\theta)$.  In the nucleon rest frame, this term describes that piece of the quark-diquark relative momentum correlation which possesses zero \emph{intrinsic} quark-diquark orbital angular momentum, \emph{i.e}.\ $L=0$ before the propagator lines are reattached to form the Faddeev wave function.  Referring to Fig.\,\ref{fig:Faddeev}, $p= P/3-p_q$ and $\cos\theta = p\cdot P/\sqrt{p^2 P^2}$.  (The amplitude is normalised such that its $U_0$ Chebyshev moment is unity at $|p|=0$.)}
\end{figure}

The quark$+$diquark structure of the nucleon is elucidated in Fig.\,\ref{figS1}, which depicts the leading component of its Faddeev amplitude: with the notation of Ref.\,\cite{Segovia:2014aza}, $s_1(|p|,\cos\theta)$, computed using the Faddeev kernel described therein.  This function describes a piece of the quark$+$scalar-diquark relative momentum correlation.
Notably, in this solution of a realistic Faddeev equation there is strong variation with respect to both arguments.  Support is concentrated in the forward direction, $\cos\theta >0$, so that alignment of $p$ and $P$ is favoured; and the amplitude peaks at $(|p|\simeq M_N/6,\cos\theta=1)$, whereat $p_q \approx P/2 \approx p_d$ and hence the \emph{natural} relative momentum is zero.  In the antiparallel direction, $\cos\theta<0$, support is concentrated at $|p|=0$, \emph{i.e}.\ $p_q \approx P/3$, $p_d \approx 2P/3$.  A realistic nucleon amplitude is evidently a complicated function; and significant structure is lost if simple interactions and/or truncations are employed in building the Faddeev kernel, \emph{e.g}.\ extant treatments of a momentum-independent quark-quark interaction -- a contact interaction -- produce a Faddeev amplitude that is also momentum independent \cite{Wilson:2011aa, Cloet:2014rja}, a result exposed as unrealistic by Fig.\,\ref{figS1} for any probe sensitive to the nucleon interior.

A nucleon (and kindred baryons) described by Fig.\,\ref{fig:Faddeev} is a Borromean bound-state, the binding within which has two contributions.  One part is expressed in the formation of tight diquark correlations.  That is augmented, however, by attraction generated by the quark exchange depicted in the shaded area of Fig.\,\ref{fig:Faddeev}.  This exchange ensures that diquark correlations within the nucleon are fully dynamical: no quark holds a special place because each one participates in all diquarks to the fullest extent allowed by its quantum numbers. The continual rearrangement of the quarks guarantees, \emph{inter} \emph{alia}, that the nucleon's dressed-quark wave function complies with Pauli statistics.

It is impossible to overstate the importance of appreciating that these fully dynamical diquark correlations are vastly different from the static, pointlike ``diquarks'' which featured in early attempts \cite{Lichtenberg:1967zz, Lichtenberg:1968zz} to understand the baryon spectrum and to explain the so-called missing resonance problem \cite{Ripani:2002ss, Burkert:2012ee, Kamano:2013iva}.  Modern diquarks are soft, Eq.\,\eqref{qqradii}; and, as we shall explain, enforce certain distinct interaction patterns for the singly- and doubly-represented valence-quarks within the proton.  On the other hand, the number of states in the spectrum of baryons obtained from the Faddeev equation in Fig.\,\ref{fig:Faddeev} \cite{Chen:2012qrS} is similar to that found in the three-constituent quark model, just as it is in today's lQCD calculations of this spectrum \cite{Edwards:2011jj}.

\smallskip

\noindent\textbf{4.$\;$Nucleon current}.
The Poincar\'e-covariant photon-nucleon interaction current is:
\begin{equation}
J_{\mu}(K,Q) = i e \,\bar{u}(P_{f})\,\left[ \gamma_{\mu} F_{1}(Q^{2}) +
\frac{\sigma_{\mu\nu}\,Q_{\nu}}{2m_{N}}\,F_{2}(Q^{2})\right] u(P_{i})\,,
\label{eq:Jnucleon}
\end{equation}
where $P_{i}$ ($P_{f}$) is the momentum of the incoming (outgoing) nucleon; $Q=P_{f} - P_{i}$, $K=(P_{i}+P_{f})/2$: for elastic scattering, $K\cdot Q=0$, $K^{2} = - m_{N}^{2} (1+\tau_{N})$, $\tau_{N} = Q^{2}/(4 m_{N}^{2})$.  The functions $F_{1,2}$ are, respectively, the Dirac and Pauli form factors: $F_1(0)$ expresses the bound-state's electric charge and $F_2(0)$, its anomalous magnetic moment, $\kappa_{N=n,p}$.  Notably, $F_2\equiv 0$ for any massless fermion \cite{Chang:2010hb}.  The Sachs electric and magnetic form factors are, respectively, $G_E = F_1 - \tau_N F_2$, $G_M = F_1+F_2$.

A nucleon described by the Faddeev equation in Fig.\,\ref{fig:Faddeev} is constituted from dressed-quarks, any two of which are always correlated as either a scalar or pseudovector diquark.  If this is a veracious description of Nature, then the presence of these correlations must be evident in numerous empirical differences between the response of the bound-state's doubly- and singly-represented quarks to any probe whose wavelength is small enough to expose the diquarks' nonpointlike character.  Associating a monopole mass with the radii in Eqs.\,\eqref{qqradii}, it becomes apparent that this wavelength corresponds to momentum transfers $Q^2 \gtrsim m_\rho^2$, where $m_\rho$ is the $\rho$-meson's mass.

In connection with electromagnetic probes, it is now possible to check these predictions following the appearance of high precision data on the neutron's electric form factor out to $Q^2=3.4\,$GeV$^2$ \cite{Riordan:2010id}.  The $G_{E}^n$ data are significant largely because they can be combined with existing empirical information on $G_{M}^n$, $G_{E,M}^p$ in order to produce a flavour separation of the proton's Dirac and Pauli form factors \cite{Cates:2011pz, Qattan:2012zf}, \emph{i.e}.\ a chart of the separate contributions of $u$- and $d$-quarks to the proton's form factors.  Supposing $s$-quark contributions are negligible, as seems the case \cite{Shanahan:2014tja}, and assuming charge symmetry, then
\begin{equation}
F_{1,2}^{u}=2F_{1,2}^{p} + F_{1,2}^{n}\,, \quad
F_{1,2}^{d}=2F_{1,2}^{n} + F_{1,2}^{p}\,.
\end{equation}
In the future, nucleon-to-resonance transition form factors might be used similarly \cite{Aznauryan:2012ba, Gothe:2009, Carman:2014}, in which event numerous new windows on baryon structure would be opened.

Evaluation of the nucleon currents is detailed in Ref.\,\cite{Segovia:2014aza} and the results we describe herein are derived from that analysis, which provides a unified description of the electromagnetic properties of the nucleon, $\Delta$-baryon and Roper resonance \cite{Segovia:2015hra}.  In what follows, it is important to note that the nucleon current can unambiguously be decomposed as follows:
\begin{equation}
\label{currentlk}
J_\mu(K,Q) = \sum_{k,l=1,\ldots,8} J_\mu^{kl}(K,Q) \,,
\end{equation}
where $k,l$, respectively, label the diquark component in the complete Faddeev wave function for the final and initial state.  For example, $J_\mu^{11}$ denotes that contribution to the current obtained when one selects for both the final and initial state a scalar diquark correlation with $L=0$ in the nucleon's rest frame, and the sum over $k,l=1,2$ expresses the complete scalar-diquark contribution.

\smallskip

\noindent\textbf{5.$\;$Verifiable predictions of diquark pairing}.
Consider the ratio of proton electric and magnetic form factors, $R_{EM}(Q^2)=\mu_p G_E(Q^2)/G_M(Q^2)$, $\mu_p=G_M(0)$.  A series of experiments \cite{Jones:1999rz, Gayou:2001qd, Gayou:2001qt, Punjabi:2005wq, Puckett:2010ac, Puckett:2011xg} has determined that $R_{EM}(Q^2)$ decreases almost linearly with $Q^2$ and might become negative for $Q^2 \gtrsim 8\,$GeV$^2$.  Our first goal is to clarify the origin of this behaviour.

\begin{figure}[t]
\centerline{%
\includegraphics[clip,width=0.4\textwidth]{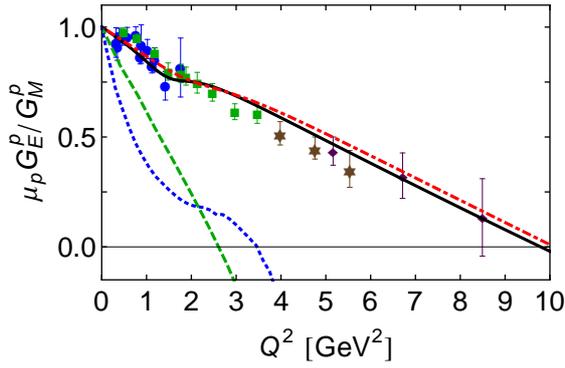}}
\caption{\label{figGEGM} Computed ratio of proton electric and magnetic form factors.  Curves:
solid (black) -- full result, determined from the complete proton Faddeev wave function and current;
dot-dashed (red) -- momentum-dependence of scalar-diquark contribution [sum over $k,l=1,2$ in Eq.\eqref{currentlk}];
dashed (green) -- momentum-dependence produced by that piece of the scalar diquark contribution to the proton's Faddeev wave function which is purely $S$-wave in the rest-frame [from $J_\mu^{11}$ in Eq.\eqref{currentlk}];
dotted (blue) -- momentum-dependence of pseudovector diquark contribution [from the sum over $k,l=3,8$ in Eq.\eqref{currentlk}].
All partial contributions have been renormalised to produce unity at $Q^2=0$.
Data:
circles (blue) \cite{Gayou:2001qt};
squares (green)  \cite{Punjabi:2005wq};
asterisks (brown) \cite{Puckett:2010ac};
and diamonds (purple) \cite{Puckett:2011xg}.
}
\end{figure}


A clear conclusion from Fig.\,\ref{figGEGM} is that pseudovector diquark correlations have little influence on the momentum dependence of $R_{EM}(Q^2)$.  Their contribution is indicated by the dotted (blue) curve, which was obtained by setting the scalar diquark component of the proton's Faddeev amplitude to zero and renormalising the result to unity at $Q^2=0$.  As apparent from the dot-dashed (red) curve, the evolution of $R_{EM}(Q^2)$ with $Q^2$ is primarily determined by the proton's scalar diquark component.  In this component, the valence $d$-quark is sequestered inside the soft scalar diquark correlation so that the only objects within the nucleon which can participate in a hard scattering event are the valence $u$-quarks.  (Any interaction with the $d$-quark attracts a $1/Q^2$ suppression because it is always locked into a correlation described by a meson-like form factor \cite{Maris:2004bp}.)

It is known from Ref.\,\cite{Cloet:2013jya} that scattering from the proton's valence $u$-quarks is responsible for the momentum dependence of $R_{EM}(Q^2)$.  However, the dashed (green) curve in Fig.\,\ref{figGEGM} reveals something more, \emph{i.e}.\ components of the nucleon associated with quark-diquark orbital angular momentum $L=1$ in the nucleon rest frame are critical in explaining the data.  Notably, the presence of such components is an inescapable consequence of the self-consistent solution of a realistic Poincar\'e-covariant Faddeev equation for the nucleon.  The visible impact on $R_{EM}(Q^2)$ is primarily driven by a marked reduction in $F_1^p$ and a lesser effect on $F_2^p$ when the $L=1$ components are neglected.  (This behaviour can be read from Figs.\,3 and 4 in Ref.\,\cite{Cloet:2008re}.)  The effect can be understood once it is recalled that a Gordon identity may be used to re-express the $\gamma_\mu$ term in Eq.\,\eqref{eq:Jnucleon} as a sum of two equally important terms, \emph{viz}.\ a convection current, as appears in the nonrelativistic case, and a spin current, which leads to a gyromagnetic ratio of two for a pointlike fermion.

\begin{figure}[t]
\centerline{%
\includegraphics[clip,width=0.40\textwidth]{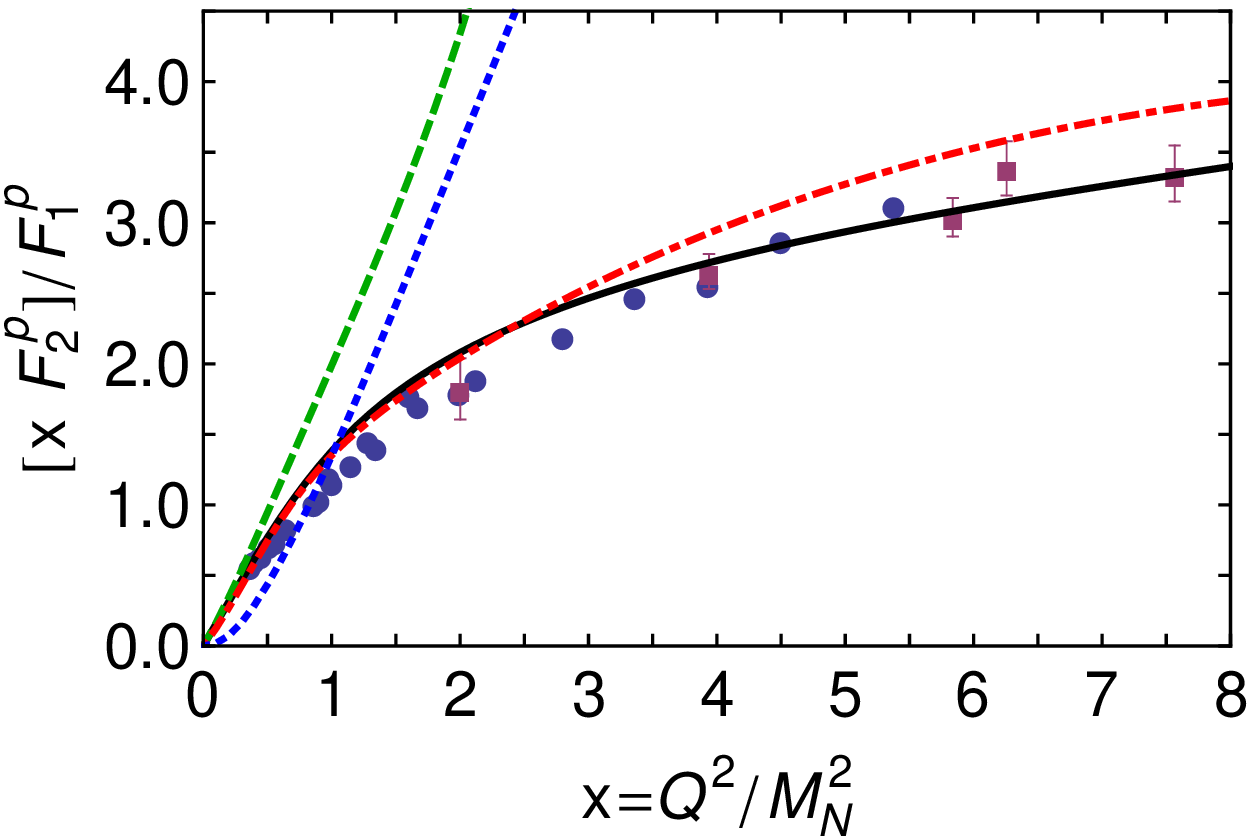}}
\vspace*{-2ex}
\centerline{%
\includegraphics[clip,width=0.40\textwidth]{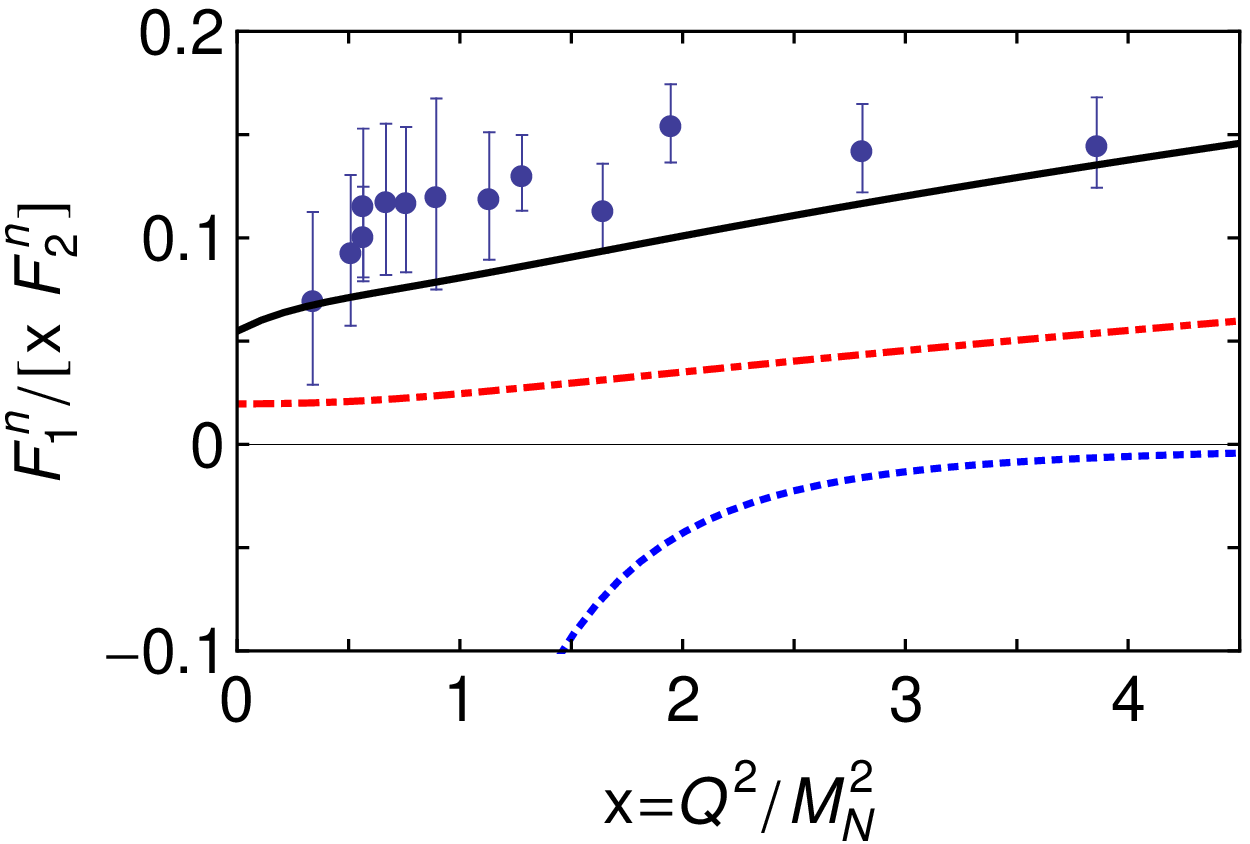}}
\caption{\label{figF1F2fla3}
\emph{Upper panel}.  Proton ratio $R_{21}(x) = x F_2(x)/F_1(x)$, $x=Q^2/M_N^2$.  Curves:
solid (black) -- full result, determined from the complete proton Faddeev wave function and current;
dot-dashed (red) -- momentum-dependence of the scalar-diquark contribution;
dashed (green) -- momentum-dependence of that component of the scalar diquark contribution to the proton's Faddeev wave function which is purely $S$-wave in the rest-frame;
dotted (blue) -- momentum-dependence of the pseudovector diquark contribution.
\emph{Lower panel}.  Neutron ratio $R_{12}^n(x) = F_1^n(x)/[x F_2^n(x)]$.  Curve legend as in the upper panel.
The data in both panels are drawn from Refs.\,\protect\cite{Zhu:2001md, Bermuth:2003qh, Warren:2003ma, Glazier:2004ny, Plaster:2005cx, Riordan:2010id, Cates:2011pz}.
}
\end{figure}

It must also be noted that the presence of diquark correlations and the use of a Poincar\'e covariant framework is \emph{insufficient} to explain the data in Fig.\,\eqref{figGEGM}.  It is possible to incorporate both but still fail in this comparison, \emph{e.g}.\ Faddeev equation studies based on a quark-quark contact interaction always generate a zero in the neighbourhood $Q^2\simeq 4\,M_N^2$ \cite{Wilson:2011aa, Cloet:2014rja}, and are thus ruled-out by the data.  As explained in Refs.\,\cite{Wilson:2011aa, Cloet:2013gva}, the flaw in those studies is the contact interaction itself, which generates a momentum-independent dressed-quark mass.  The existence and location of a zero in $R_{EM}(Q^2)$  are a measure of nonperturbative features of the quark-quark interaction, with particular sensitivity to the running of the dressed-quark mass \cite{Cloet:2013gva}.

It is natural now to consider the proton ratio: $R_{21}(x) = x F_2(x)/F_1(x)$, $x=Q^2/M_N^2$, drawn in the upper panel of Fig.\,\ref{figF1F2fla3}.  As with $R_{EM}$, the momentum dependence of $R_{21}(x)$ is principally determined by the scalar diquark component of the proton.  Moreover, the rest-frame $L=1$ terms are again seen to be critical in explaining the data: the behaviour of the dashed (green) curve highlights the impact of omitting these components.

These remarks concerning Faddeev wave function components with quark-diquark orbital angular momentum $L\neq 0$ in the nucleon rest frame are consistent with the relativistic constituent-quark model study of Ref.\,\cite{Miller:2002qb} and the analysis of nucleon spin structure in Ref.\,\cite{Roberts:2013mja}.  Both explain that at energies accessible now and for the foreseeable future, the nucleon is described by a complex wave function that can be characterised as possessing significant quark orbital angular momentum.  This being the case, then helicity conservation can neither be a good approximation in the analysis of extant measurements nor a reliable guide to the interpretation of anticipated data.  The same effect is manifest in analyses of the $N\to\Delta$ transition \cite{Eichmann:2011aa, Segovia:2014aza}.

The lower panel of Fig.\,\ref{figF1F2fla3} displays an analogous ratio for the neutron: $R_{12}^n(x) = F_1^n(x)/[x F_2^n(x)]$.  Here the curve obtained in the absence of pseudovector diquarks does not resemble the data, despite the fact that both the scalar-diquark-only and pseudovector-diquark-only curves are finite at $x=0$.  Apparently, something more than orbital angular momentum and a running quark mass is important in understanding and explaining the behaviour of nucleon electromagnetic form factors; and whatever it is must distinguish between isospin partners.  This could have been anticipated from Ref.\,\cite{Cloet:2008re} through a comparison of Figs.\,6 and 13 therein: whilst so-called precocious scaling was evident in $R_{21}(x)$, this was not the case for $R_{21}^n(x)$.  The additional feature, of course, is the presence of scalar and pseudovector diquark correlations, which have different impacts on the doubly and singly represented valence-quarks.


\begin{figure}[!t]
\centerline{\includegraphics[clip,width=0.40\textwidth]{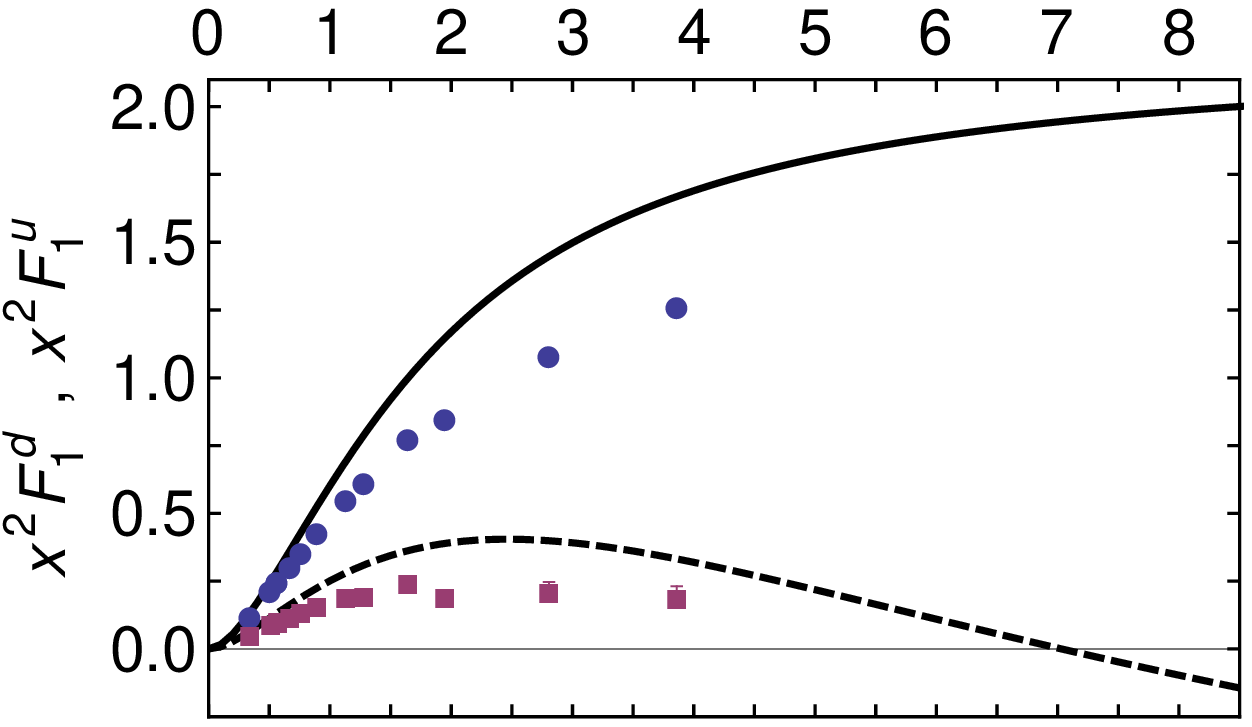}}
\vspace*{-1.6ex}

\centerline{\includegraphics[clip,width=0.40\textwidth]{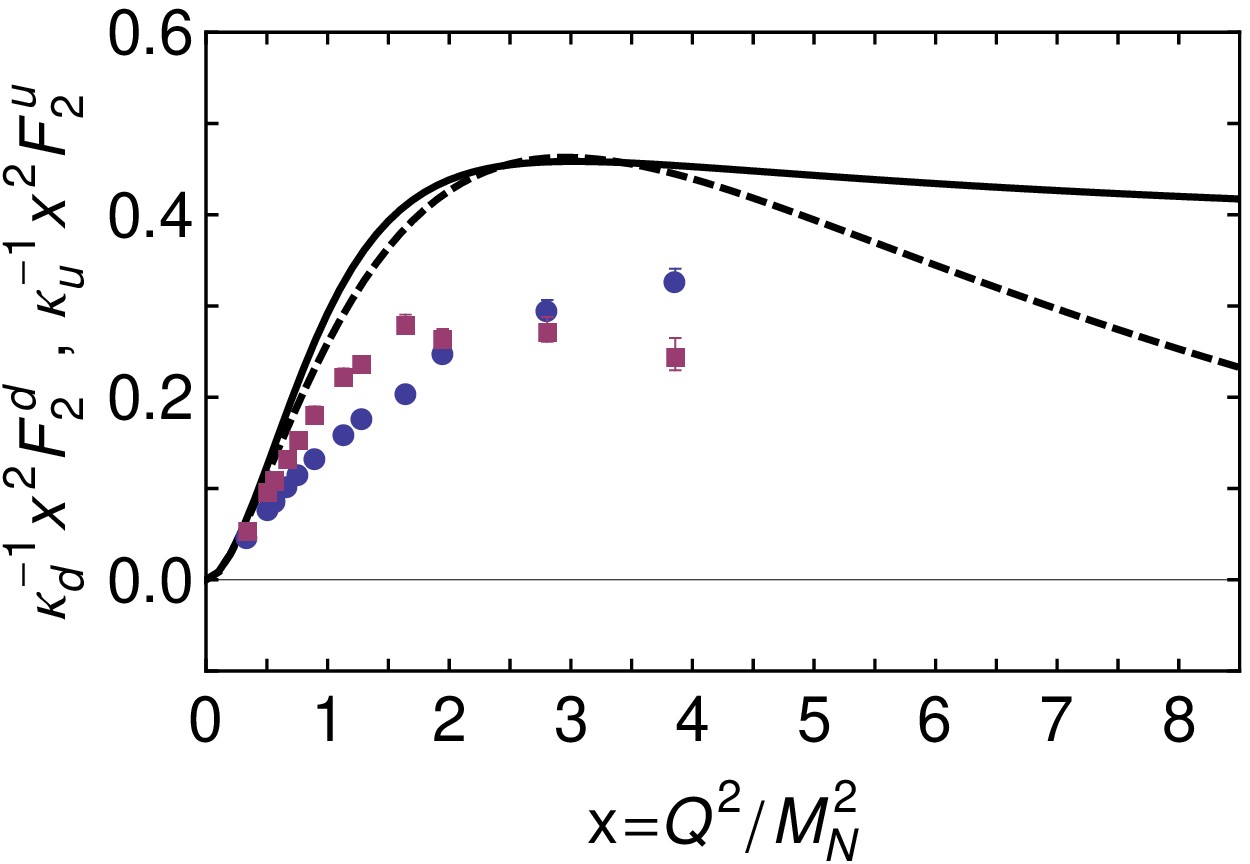}}
\caption{\label{fig:F1F2fla1} \emph{Upper panel}.  Flavour separation of the proton's Dirac form factor as a function of $x=Q^2/M_N^2$.  Curves: solid -- $u$-quark; and dashed $d$-quark contribution.  Data: circles -- $u$-quark; and squares -- $d$-quark.
\emph{Lower panel}.  Same for Pauli form factor.
Data: Refs.\,\protect\cite{Zhu:2001md, Bermuth:2003qh, Warren:2003ma, Glazier:2004ny, Plaster:2005cx, Riordan:2010id, Cates:2011pz}.
}
\end{figure}

Figure\,\ref{fig:F1F2fla1} displays the proton's flavour separated Dirac and Pauli form factors.  The salient features of the data are: the $d$-quark contribution to $F_1^p$ is far smaller than the $u$-quark contribution; $F_2^d/\kappa_d>F_2^u/\kappa_u$ on $x<2$ but this ordering is reversed on $x>2$; and in both cases the $d$-quark contribution falls dramatically on $x>3$ whereas the $u$-quark contribution remains roughly constant.   Our calculations are in semi-quantitative agreement with the empirical data.  They reproduce the qualitative behaviour and also predict a zero in $F_1^d$ at $x\simeq 7$.

It is natural to seek an explanation for the pattern of behaviour in Fig.\,\ref{fig:F1F2fla1}.  We have emphasised that the proton contains scalar and pseudovector diquark correlations.  The dominant piece of its Faddeev wave function is $u[ud]$; namely, a $u$-quark in tandem with a $[ud]$ scalar correlation, which produces 62\% of the proton's normalisation \cite{Cloet:2007piS}.  If this were the sole component, then photon--$d$-quark interactions within the proton would receive a $1/x$ suppression on $x>1$, because the $d$-quark is sequestered in a soft correlation, whereas a spectator $u$-quark is always available to participate in a hard interaction.  At large $x=Q^2/M_N^2$, therefore, scalar diquark dominance leads one to expect $F^d \sim F^u/x$.  Available data are consistent with this prediction but measurements at $x>4$ are necessary for confirmation.  Furthermore, as first remarked in Refs.\,\cite{Roberts:2010huS, Cloet:2011qu}, scalar diquark correlations cannot be the entire explanation because they alone cannot produce a zero in $F_1^d$.

\begin{figure}[t]
\centerline{\includegraphics[clip,width=0.40\textwidth]{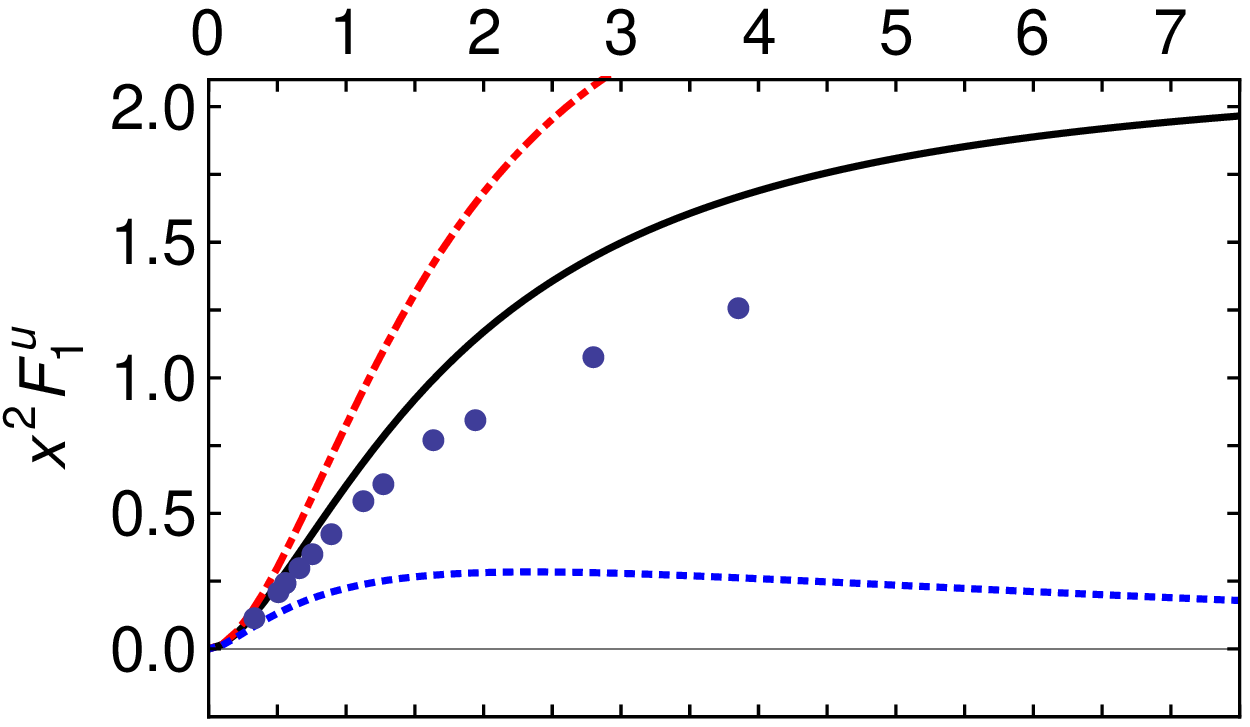}}
\vspace*{-1.6ex}

\centerline{\includegraphics[clip,width=0.40\textwidth]{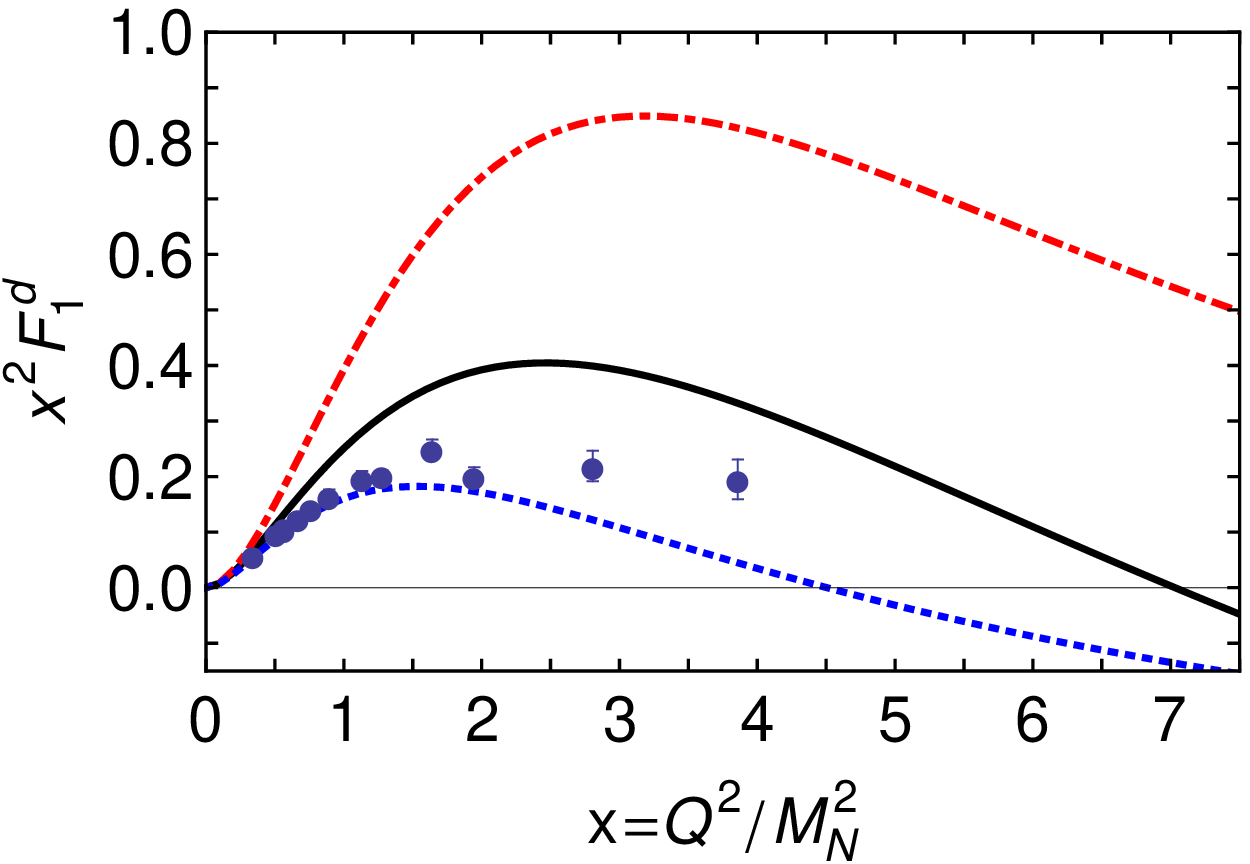}}
\caption{\label{fig:F1uF1d} \emph{Upper panel}.  $u$-quark contribution to the proton's Dirac form factor as a function of $x=Q^2/M_N^2$.  Curves: solid (black) -- complete contribution; dot-dashed (red) -- scalar-diquark contribution;
dotted (blue) -- pseudovector diquark contribution.
\emph{Lower panel}.  $d$-quark contribution to the proton's Dirac form factor.  Curve legend same as upper panel.
Data: Refs.\,\protect\cite{Zhu:2001md, Bermuth:2003qh, Warren:2003ma, Glazier:2004ny, Plaster:2005cx, Riordan:2010id, Cates:2011pz}.
}
\end{figure}

Consider the images in Fig.\,\ref{fig:F1uF1d}, which expose the relative strength of scalar and pseudovector correlations in the flavour separated form factors.  The upper panel shows that whilst the scalar diquark component of the proton is the dominant determining feature of $F_1^u$, \emph{i.e}.\ in connection with the doubly represented valence-quark, the pseudovector component nevertheless plays a measurable role.

In the case of $F_1^d$ (lower panel, Fig.\,\ref{fig:F1uF1d}) the pseudovector correlation provides the leading contribution.  The proton's pseudovector component appears in two combinations: $u\{ud\}$ and $d\{uu\}$.  The latter involves a hard $d$-quark and is twice as probable as the former (isospin Clebsch-Gordon algebra).  The presence of pseudovector diquarks in the proton therefore guarantees that valence $d$-quarks will always be available to participate in a hard scattering event.
$F_1^d$ possesses a zero because so does each of its separated contributions.  (This is evident in Ref.\,\cite{Cloet:2008re}, discussion of Fig.\,3, lower-right panel.)  The location of the predicted zero therefore depends on the strength of interference with the scalar diquark part of the proton.
Hence, like the ratios of valence-quark parton distribution functions at large Bjorken-$x$ \cite{Roberts:2013mja, Parno:2014xzb}, the location of the zero in $F_1^d$ is a measure of the relative probability of finding pseudovector and scalar diquarks in the proton:  with all other things held equal, the zero moves toward $x=0$ as the probability of finding a pseudovector diquark within the proton increases.  Empirical verification of a zero in $F_1^d$ would be definitive evidence that the ``precocious scaling'' of $R_{21}(x)$ is accidental, existing only on a narrow domain because of fortuitous cancellations amongst the many scattering diagrams involved in expressing the current of a proton comprised from tight quark-quark correlations.

\begin{figure}[t]
\centerline{\includegraphics[clip,width=0.40\textwidth]{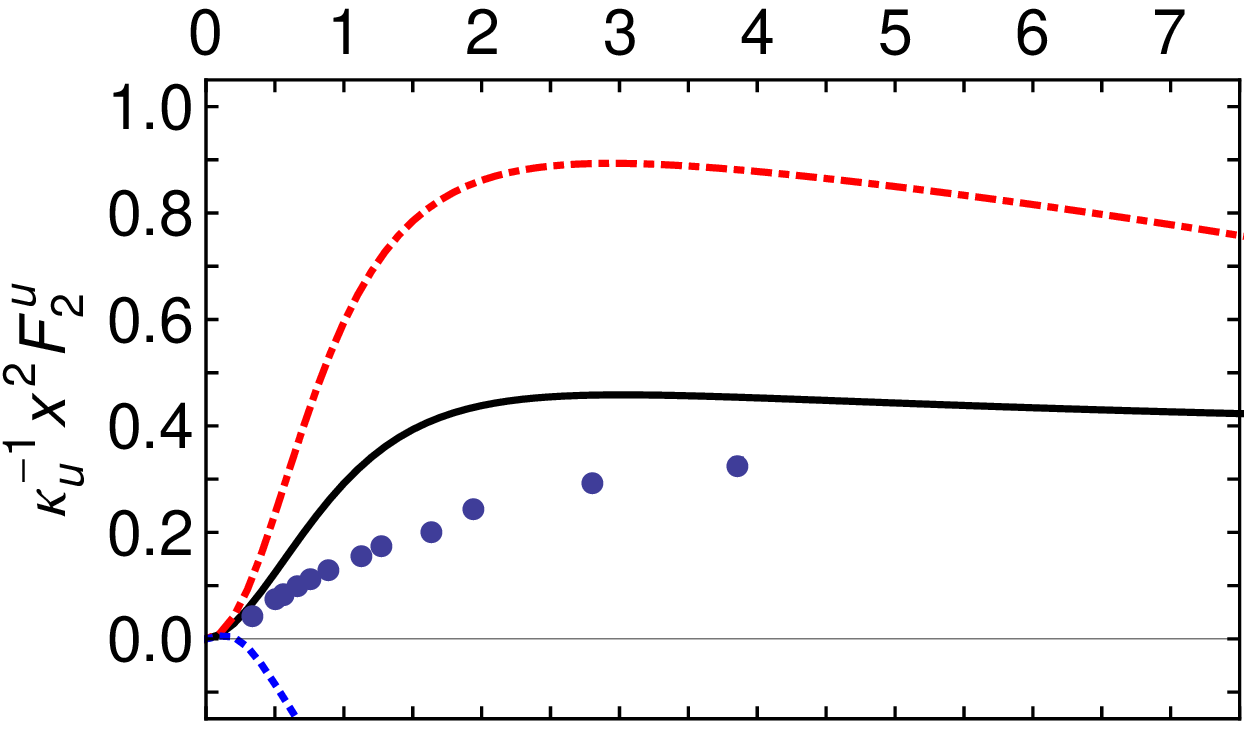}}
\vspace*{-1.6ex}

\centerline{\includegraphics[clip,width=0.40\textwidth]{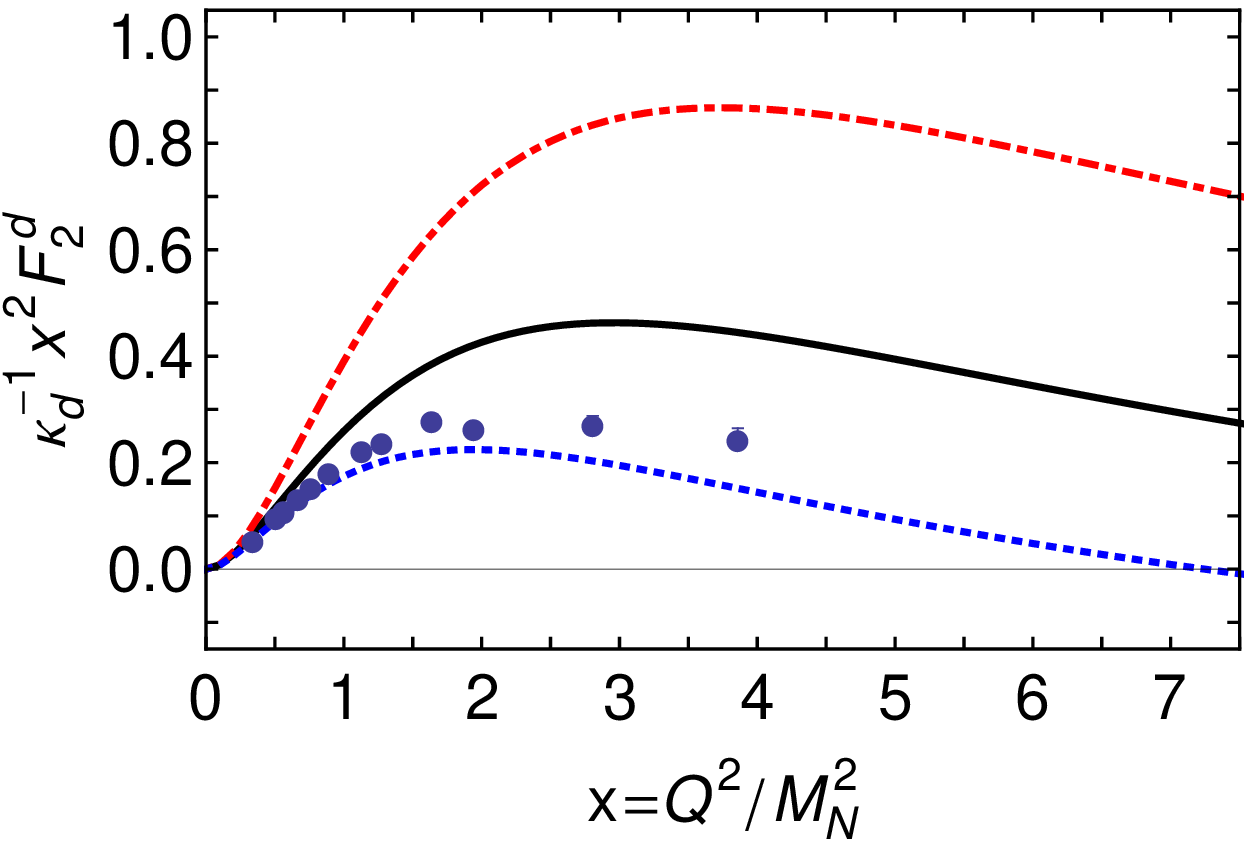}}
\caption{\label{fig:F1F2fla2}
\emph{Upper panel}.  $u$-quark contribution to the proton's Pauli form factor as a function of $x=Q^2/M_N^2$.  Curves: solid (black) -- complete contribution; dot-dashed (red) -- scalar-diquark contribution;
dotted (blue) -- pseudovector diquark contribution.
\emph{Lower panel}.  $d$-quark contribution to the proton's Pauli form factor.  Curve legend same as upper panel.
Data: Refs.\,\protect\cite{Zhu:2001md, Bermuth:2003qh, Warren:2003ma, Glazier:2004ny, Plaster:2005cx, Riordan:2010id, Cates:2011pz}.
}
\end{figure}

In Fig.\,\ref{fig:F1F2fla2} we draw analogous figures for the proton's flavour-separated Pauli form factor.  Plainly, $F_2^u$ is far more sensitive to interference between scalar and pseudovector diquark correlations than $F_1^u$.  On the other hand, $F_{1,2}^d$ exhibit similar patterns of interplay between scalar and pseudovector diquarks.

The information contained in Figs.\,\ref{fig:F1F2fla1} -- \ref{fig:F1F2fla2} provides clear evidence in support of the notion that many features in the measured behaviour of nucleon electromagnetic form factors are primarily determined by the presence of strong diquark correlations in the nucleon.  Importantly, whilst inclusion of a ``pion cloud'' can potentially improve quantitative agreement with data, it does not qualitatively affect the salient features of the form factors \cite{Cloet:2012cy, Cloet:2014rja}.

\smallskip

\noindent\textbf{6.$\;$Summary}.
We explained how the emergent phenomenon of dynamical chiral symmetry breaking ensures that Poincar\'e covariant analyses of the three valence-quark scattering problem in continuum quantum field theory yield a picture of the nucleon as a Borromean bound-state, in which binding arises primarily through the sum of two separate contributions.  One involves aspects of the non-Abelian character of QCD that are expressed in the strong running coupling and generate tight, dynamical colour-antitriplet quark-quark correlations in the scalar-isoscalar and pseudovector-isotriplet channels.  This attraction is magnified by quark exchange associated with diquark breakup and reformation, which is required in order to ensure that each valence-quark participates in all diquark correlations to the complete extent allowed by its quantum numbers.

Combining these effects, one arrives at a properly antisymmetrised Faddeev wave function for the nucleon and is positioned to compute a wide range of observables.  Capitalising on this, we illustrated and emphasised that numerous empirical consequences derive from: Poincar\'e covariance, which demands the presence of dressed-quark orbital angular momentum in the nucleon; the behaviour of the strong running coupling as expressed, for instance, in the momentum-dependence of the dressed-quark mass; and the existence of strong electromagnetically-active scalar and pseudovector diquark correlations within the nucleon, which ensure marked differences between properties associated with doubly- and singly-represented valence-quarks.  Planned experiments are therefore capable of validating the proposed picture of the nucleon and placing tight constraints, \emph{e.g}.\ on the rate at which dressed-quarks shed their clothing and transform into partons, and the relative probability of finding scalar and pseudovector diquarks within the nucleon.

\smallskip

\noindent\textbf{Acknowledgements}.
We are grateful for insightful comments from
D.~Binosi,
I.\,C.~Clo\"et,
R.~Gothe,
T.-S.\,H.~Lee,
V.~Mokeev,
J.~Papavassiliou,
S.-X.~Qin,
T.~Sato
and S.-S.~Xu.
J.\,Segovia acknowledges financial support from a postdoctoral IUFFyM contract
at \emph{Universidad de Salamanca}, Spain.
Work otherwise supported by U.S.\ Department of Energy, Office of Science, Office of Nuclear Physics, under contract no.~DE-AC02-06CH11357.

\smallskip



\begin{thebibliography}{92}
\expandafter\ifx\csname natexlab\endcsname\relax\def\natexlab#1{#1}\fi
\providecommand{\bibinfo}[2]{#2}
\ifx\xfnm\relax \def\xfnm[#1]{\unskip,\space#1}\fi
\bibitem[{Olive et~al.(2014)}]{Agashe:2014kda}
\bibinfo{author}{K.~A. Olive}, et~al., \bibinfo{journal}{Chin. Phys. C}
  \bibinfo{volume}{38} (\bibinfo{year}{2014}) \bibinfo{pages}{090001}.
\bibitem[{Dirac(1949)}]{Dirac:1949cp}
\bibinfo{author}{P.~A.~M. Dirac}, \bibinfo{journal}{Rev. Mod. Phys.}
  \bibinfo{volume}{21} (\bibinfo{year}{1949}) \bibinfo{pages}{392--399}.
\bibitem[{Keister and Polyzou(1991)}]{Keister:1991sb}
\bibinfo{author}{B.~D. Keister}, \bibinfo{author}{W.~N. Polyzou},
  \bibinfo{journal}{Adv. Nucl. Phys.} \bibinfo{volume}{20}
  (\bibinfo{year}{1991}) \bibinfo{pages}{225--479}.
\bibitem[{Coester(1992)}]{Coester:1992cg}
\bibinfo{author}{F.~Coester}, \bibinfo{journal}{Prog. Part. Nucl. Phys.}
  \bibinfo{volume}{29} (\bibinfo{year}{1992}) \bibinfo{pages}{1--32}.
\bibitem[{Brodsky et~al.(1998)Brodsky, Pauli, and Pinsky}]{Brodsky:1997de}
\bibinfo{author}{S.~J. Brodsky}, \bibinfo{author}{H.-C. Pauli},
  \bibinfo{author}{S.~S. Pinsky}, \bibinfo{journal}{Phys. Rept.}
  \bibinfo{volume}{301} (\bibinfo{year}{1998}) \bibinfo{pages}{299--486}.
\bibitem[{Bissey et~al.(2007)}]{Bissey:2006bz}
\bibinfo{author}{F.~Bissey}, et~al., \bibinfo{journal}{Phys. Rev. D}
  \bibinfo{volume}{76} (\bibinfo{year}{2007}) \bibinfo{pages}{114512}.
\bibitem[{Bissey et~al.(2009)Bissey, Signal, and Leinweber}]{Bissey:2009gw}
\bibinfo{author}{F.~Bissey}, \bibinfo{author}{A.~Signal},
  \bibinfo{author}{D.~Leinweber}, \bibinfo{journal}{Phys. Rev. D}
  \bibinfo{volume}{80} (\bibinfo{year}{2009}) \bibinfo{pages}{114506}.
\bibitem[{Politzer(2005)}]{Politzer:2005kc}
\bibinfo{author}{H.~Politzer}, \bibinfo{journal}{Proc. Nat. Acad. Sci.}
  \bibinfo{volume}{102} (\bibinfo{year}{2005}) \bibinfo{pages}{7789--7793}.
\bibitem[{Gross(2005)}]{Gross:2005kv}
\bibinfo{author}{D.~Gross}, \bibinfo{journal}{Proc. Nat. Acad. Sci.}
  \bibinfo{volume}{102} (\bibinfo{year}{2005}) \bibinfo{pages}{9099--9108}.
\bibitem[{Wilczek(2005)}]{Wilczek:2005az}
\bibinfo{author}{F.~Wilczek}, \bibinfo{journal}{Proc. Nat. Acad. Sci.}
  \bibinfo{volume}{102} (\bibinfo{year}{2005}) \bibinfo{pages}{8403--8413}.
\bibitem[{Freedman et~al.(2012)}]{national2012NuclearS}
\bibinfo{author}{S.~J. Freedman}, et~al., \bibinfo{title}{\emph{Nuclear
  Physics: Exploring the Heart of Matter}}, \bibinfo{publisher}{National
  Academies Press, Washington D.C.}, \bibinfo{year}{2012}.
\bibitem[{Brodsky et~al.(p ph)Brodsky, Deshpande, Gao, McKeown, Meyer, Meziani,
  Milner, Qiu, Richards, and Roberts}]{Brodsky:2015aia}
\bibinfo{author}{S.~J. Brodsky}, \bibinfo{author}{A.~L. Deshpande},
  \bibinfo{author}{H.~Gao}, \bibinfo{author}{R.~D. McKeown},
  \bibinfo{author}{C.~A. Meyer}, \bibinfo{author}{Z.-E. Meziani},
  \bibinfo{author}{R.~G. Milner}, \bibinfo{author}{J.-W. Qiu},
  \bibinfo{author}{D.~G. Richards}, \bibinfo{author}{C.~D. Roberts}
  (\bibinfo{year}{aXiv:1502.05728 [hep-ph]}). \bibinfo{note}{{\emph{QCD and
  Hadron Physics}}}.
\bibitem[{Gribov(1999)}]{Gribov:1999ui}
\bibinfo{author}{V.~N. Gribov}, \bibinfo{journal}{Eur. Phys. J. C}
  \bibinfo{volume}{10} (\bibinfo{year}{1999}) \bibinfo{pages}{91--105}.
\bibitem[{Munczek and Nemirovsky(1983)}]{Munczek:1983dx}
\bibinfo{author}{H.~J. Munczek}, \bibinfo{author}{A.~M. Nemirovsky},
  \bibinfo{journal}{Phys. Rev. D} \bibinfo{volume}{28} (\bibinfo{year}{1983})
  \bibinfo{pages}{181--186}.
\bibitem[{Stingl(1984)}]{Stingl:1983pt}
\bibinfo{author}{M.~Stingl}, \bibinfo{journal}{Phys. Rev. D}
  \bibinfo{volume}{29} (\bibinfo{year}{1984}) \bibinfo{pages}{2105}.
\bibitem[{Cahill(1989)}]{Cahill:1988zi}
\bibinfo{author}{R.~T. Cahill}, \bibinfo{journal}{Austral. J. Phys.}
  \bibinfo{volume}{42} (\bibinfo{year}{1989}) \bibinfo{pages}{171--186}.
\bibitem[{Roberts et~al.(1992)Roberts, Williams, and Krein}]{Krein:1990sf}
\bibinfo{author}{C.~D. Roberts}, \bibinfo{author}{A.~G. Williams},
  \bibinfo{author}{G.~Krein}, \bibinfo{journal}{Int. J. Mod. Phys. A}
  \bibinfo{volume}{7} (\bibinfo{year}{1992}) \bibinfo{pages}{5607--5624}.
\bibitem[{Maris et~al.(1998)Maris, Roberts, and Tandy}]{Maris:1997hd}
\bibinfo{author}{P.~Maris}, \bibinfo{author}{C.~D. Roberts},
  \bibinfo{author}{P.~C. Tandy}, \bibinfo{journal}{Phys. Lett. B}
  \bibinfo{volume}{420} (\bibinfo{year}{1998}) \bibinfo{pages}{267--273}.
\bibitem[{Bali et~al.(2005)}]{Bali:2005fuS}
\bibinfo{author}{G.~S. Bali}, et~al., \bibinfo{journal}{Phys. Rev. D}
  \bibinfo{volume}{71} (\bibinfo{year}{2005}) \bibinfo{pages}{114513}.
\bibitem[{Prkacin et~al.(2006)}]{Prkacin:2005dc}
\bibinfo{author}{Z.~Prkacin}, et~al., \bibinfo{journal}{PoS}
  \bibinfo{volume}{LAT2005} (\bibinfo{year}{2006}) \bibinfo{pages}{308}.
\bibitem[{Cahill et~al.(1987)Cahill, Roberts, and Praschifka}]{Cahill:1987qr}
\bibinfo{author}{R.~T. Cahill}, \bibinfo{author}{C.~D. Roberts},
  \bibinfo{author}{J.~Praschifka}, \bibinfo{journal}{Phys. Rev. D}
  \bibinfo{volume}{36} (\bibinfo{year}{1987}) \bibinfo{pages}{2804}.
\bibitem[{Cahill et~al.(1989)Cahill, Roberts, and Praschifka}]{Cahill:1988dx}
\bibinfo{author}{R.~T. Cahill}, \bibinfo{author}{C.~D. Roberts},
  \bibinfo{author}{J.~Praschifka}, \bibinfo{journal}{Austral. J. Phys.}
  \bibinfo{volume}{42} (\bibinfo{year}{1989}) \bibinfo{pages}{129--145}.
\bibitem[{Bender et~al.(1996)Bender, Roberts, and von Smekal}]{Bender:1996bb}
\bibinfo{author}{A.~Bender}, \bibinfo{author}{C.~D. Roberts},
  \bibinfo{author}{L.~von Smekal}, \bibinfo{journal}{Phys. Lett. B}
  \bibinfo{volume}{380} (\bibinfo{year}{1996}) \bibinfo{pages}{7--12}.
\bibitem[{Oettel et~al.(1998)Oettel, Hellstern, Alkofer, and
  Reinhardt}]{Oettel:1998bk}
\bibinfo{author}{M.~Oettel}, \bibinfo{author}{G.~Hellstern},
  \bibinfo{author}{R.~Alkofer}, \bibinfo{author}{H.~Reinhardt},
  \bibinfo{journal}{Phys. Rev. C} \bibinfo{volume}{58} (\bibinfo{year}{1998})
  \bibinfo{pages}{2459--2477}.
\bibitem[{Bloch et~al.(1999)Bloch, Roberts, and Schmidt}]{Bloch:1999vk}
\bibinfo{author}{J.~C.~R. Bloch}, \bibinfo{author}{C.~D. Roberts},
  \bibinfo{author}{S.~M. Schmidt}, \bibinfo{journal}{Phys. Rev. C}
  \bibinfo{volume}{60} (\bibinfo{year}{1999}) \bibinfo{pages}{065208}.
\bibitem[{Maris(2002)}]{Maris:2002yu}
\bibinfo{author}{P.~Maris}, \bibinfo{journal}{Few Body Syst.}
  \bibinfo{volume}{32} (\bibinfo{year}{2002}) \bibinfo{pages}{41--52}.
\bibitem[{Bender et~al.(2002)Bender, Detmold, Roberts, and
  Thomas}]{Bender:2002as}
\bibinfo{author}{A.~Bender}, \bibinfo{author}{W.~Detmold},
  \bibinfo{author}{C.~D. Roberts}, \bibinfo{author}{A.~W. Thomas},
  \bibinfo{journal}{Phys. Rev. C} \bibinfo{volume}{65} (\bibinfo{year}{2002})
  \bibinfo{pages}{065203}.
\bibitem[{Bhagwat et~al.(2004)Bhagwat, H{\"o}ll, Krassnigg, Roberts, and
  Tandy}]{Bhagwat:2004hn}
\bibinfo{author}{M.~S. Bhagwat}, \bibinfo{author}{A.~H{\"o}ll},
  \bibinfo{author}{A.~Krassnigg}, \bibinfo{author}{C.~D. Roberts},
  \bibinfo{author}{P.~C. Tandy}, \bibinfo{journal}{Phys. Rev. C}
  \bibinfo{volume}{70} (\bibinfo{year}{2004}) \bibinfo{pages}{035205}.
\bibitem[{Clo{\"e}t et~al.(2009)Clo{\"e}t, Eichmann, El-Bennich, Kl{\"a}hn, and
  Roberts}]{Cloet:2008re}
\bibinfo{author}{I.~C. Clo{\"e}t}, \bibinfo{author}{G.~Eichmann},
  \bibinfo{author}{B.~El-Bennich}, \bibinfo{author}{T.~Kl{\"a}hn},
  \bibinfo{author}{C.~D. Roberts}, \bibinfo{journal}{Few Body Syst.}
  \bibinfo{volume}{46} (\bibinfo{year}{2009}) \bibinfo{pages}{1--36}.
\bibitem[{Eichmann et~al.(2009)Eichmann, Clo{\"e}t, Alkofer, Krassnigg, and
  Roberts}]{Eichmann:2008ef}
\bibinfo{author}{G.~Eichmann}, \bibinfo{author}{I.~C. Clo{\"e}t},
  \bibinfo{author}{R.~Alkofer}, \bibinfo{author}{A.~Krassnigg},
  \bibinfo{author}{C.~D. Roberts}, \bibinfo{journal}{Phys. Rev. C}
  \bibinfo{volume}{79} (\bibinfo{year}{2009}) \bibinfo{pages}{012202(R)}.
\bibitem[{Eichmann et~al.(2010)Eichmann, Alkofer, Krassnigg, and
  Nicmorus}]{Eichmann:2009qa}
\bibinfo{author}{G.~Eichmann}, \bibinfo{author}{R.~Alkofer},
  \bibinfo{author}{A.~Krassnigg}, \bibinfo{author}{D.~Nicmorus},
  \bibinfo{journal}{Phys. Rev. Lett.} \bibinfo{volume}{104}
  (\bibinfo{year}{2010}) \bibinfo{pages}{201601}.
\bibitem[{Roberts et~al.(2011)Roberts, Chang, Clo{\"e}t, and
  Roberts}]{Roberts:2011cf}
\bibinfo{author}{H.~L.~L. Roberts}, \bibinfo{author}{L.~Chang},
  \bibinfo{author}{I.~C. Clo{\"e}t}, \bibinfo{author}{C.~D. Roberts},
  \bibinfo{journal}{Few Body Syst.} \bibinfo{volume}{51} (\bibinfo{year}{2011})
  \bibinfo{pages}{1--25}.
\bibitem[{Eichmann and Nicmorus(2012)}]{Eichmann:2011aa}
\bibinfo{author}{G.~Eichmann}, \bibinfo{author}{D.~Nicmorus},
  \bibinfo{journal}{Phys. Rev. D} \bibinfo{volume}{85} (\bibinfo{year}{2012})
  \bibinfo{pages}{093004}.
\bibitem[{Segovia et~al.(2014)Segovia, Clo{\"e}t, Roberts, and
  Schmidt}]{Segovia:2014aza}
\bibinfo{author}{J.~Segovia}, \bibinfo{author}{I.~C. Clo{\"e}t},
  \bibinfo{author}{C.~D. Roberts}, \bibinfo{author}{S.~M. Schmidt},
  \bibinfo{journal}{Few Body Syst.} \bibinfo{volume}{55} (\bibinfo{year}{2014})
  \bibinfo{pages}{1185--1222}.
\bibitem[{Segovia et~al.(l th)Segovia, El-Bennich, Rojas, Cloet, Roberts
  et~al.}]{Segovia:2015hra}
\bibinfo{author}{J.~Segovia}, \bibinfo{author}{B.~El-Bennich},
  \bibinfo{author}{E.~Rojas}, \bibinfo{author}{I.~C. Cloet},
  \bibinfo{author}{C.~D. Roberts}, et~al.  (\bibinfo{year}{arXiv:1504.04386
  [nucl-th]}). \bibinfo{note}{{\emph{Completing the picture of the Roper
  resonance}}}.
\bibitem[{Alexandrou et~al.(2006)Alexandrou, de~Forcrand, and
  Lucini}]{Alexandrou:2006cq}
\bibinfo{author}{C.~Alexandrou}, \bibinfo{author}{{\mbox{Ph}}.~de~Forcrand},
  \bibinfo{author}{B.~Lucini}, \bibinfo{journal}{Phys. Rev. Lett.}
  \bibinfo{volume}{97} (\bibinfo{year}{2006}) \bibinfo{pages}{222002}.
\bibitem[{Babich et~al.(2007)Babich, Garron, Hoelbling, Howard, Lellouch
  et~al.}]{Babich:2007ah}
\bibinfo{author}{R.~Babich}, \bibinfo{author}{N.~Garron},
  \bibinfo{author}{C.~Hoelbling}, \bibinfo{author}{J.~Howard},
  \bibinfo{author}{L.~Lellouch}, et~al., \bibinfo{journal}{Phys. Rev. D}
  \bibinfo{volume}{76} (\bibinfo{year}{2007}) \bibinfo{pages}{074021}.
\bibitem[{Munczek(1995)}]{Munczek:1994zz}
\bibinfo{author}{H.~J. Munczek}, \bibinfo{journal}{Phys. Rev. D}
  \bibinfo{volume}{52} (\bibinfo{year}{1995}) \bibinfo{pages}{4736--4740}.
\bibitem[{Maris(2004)}]{Maris:2004bp}
\bibinfo{author}{P.~Maris}, \bibinfo{journal}{Few Body Syst.}
  \bibinfo{volume}{35} (\bibinfo{year}{2004}) \bibinfo{pages}{117--127}.
\bibitem[{Roberts et~al.(2011)}]{Roberts:2011wyS}
\bibinfo{author}{H.~L.~L. Roberts}, et~al., \bibinfo{journal}{Phys. Rev. C}
  \bibinfo{volume}{83} (\bibinfo{year}{2011}) \bibinfo{pages}{065206}.
\bibitem[{Pauli(1957)}]{Pauli:1957}
\bibinfo{author}{W.~Pauli}, \bibinfo{journal}{Nuovo Cim.} \bibinfo{volume}{6}
  (\bibinfo{year}{1957}) \bibinfo{pages}{204--215}.
\bibitem[{G{\"u}rsey(1958)}]{Gursey:1958}
\bibinfo{author}{F.~G{\"u}rsey}, \bibinfo{journal}{Nuovo Cim.}
  \bibinfo{volume}{7} (\bibinfo{year}{1958}) \bibinfo{pages}{411--415}.
\bibitem[{Aguilar et~al.(2008)Aguilar, Binosi, and
  Papavassiliou}]{Aguilar:2008xm}
\bibinfo{author}{A.~Aguilar}, \bibinfo{author}{D.~Binosi},
  \bibinfo{author}{J.~Papavassiliou}, \bibinfo{journal}{Phys. Rev. D}
  \bibinfo{volume}{78} (\bibinfo{year}{2008}) \bibinfo{pages}{025010}.
\bibitem[{Aguilar and Papavassiliou(2010)}]{Aguilar:2009ke}
\bibinfo{author}{A.~C. Aguilar}, \bibinfo{author}{J.~Papavassiliou},
  \bibinfo{journal}{Phys. Rev. D} \bibinfo{volume}{81} (\bibinfo{year}{2010})
  \bibinfo{pages}{034003}.
\bibitem[{Aguilar et~al.(2009)Aguilar, Binosi, Papavassiliou, and
  Rodr{\'i}guez-Quintero}]{Aguilar:2009nf}
\bibinfo{author}{A.~Aguilar}, \bibinfo{author}{D.~Binosi},
  \bibinfo{author}{J.~Papavassiliou},
  \bibinfo{author}{J.~Rodr{\'i}guez-Quintero}, \bibinfo{journal}{Phys. Rev. D}
  \bibinfo{volume}{80} (\bibinfo{year}{2009}) \bibinfo{pages}{085018}.
\bibitem[{Maas(2013)}]{Maas:2011se}
\bibinfo{author}{A.~Maas}, \bibinfo{journal}{Phys. Rept.} \bibinfo{volume}{524}
  (\bibinfo{year}{2013}) \bibinfo{pages}{203--300}.
\bibitem[{Boucaud et~al.(2012)Boucaud, Leroy, Le-Yaouanc, Micheli, Pene, and
  Rodr{\'i}guez-Quintero}]{Boucaud:2011ug}
\bibinfo{author}{P.~Boucaud}, \bibinfo{author}{J.~P. Leroy},
  \bibinfo{author}{A.~Le-Yaouanc}, \bibinfo{author}{J.~Micheli},
  \bibinfo{author}{O.~Pene}, \bibinfo{author}{J.~Rodr{\'i}guez-Quintero},
  \bibinfo{journal}{Few Body Syst.} \bibinfo{volume}{53} (\bibinfo{year}{2012})
  \bibinfo{pages}{387--436}.
\bibitem[{Pennington and Wilson(2011)}]{Pennington:2011xs}
\bibinfo{author}{M.~R. Pennington}, \bibinfo{author}{D.~J. Wilson},
  \bibinfo{journal}{Phys. Rev. D} \bibinfo{volume}{84} (\bibinfo{year}{2011})
  \bibinfo{pages}{119901}.
\bibitem[{Binosi et~al.(2012)Binosi, Ib{\'a}{\~n}ez, and
  Papavassiliou}]{Binosi:2012sj}
\bibinfo{author}{D.~Binosi}, \bibinfo{author}{D.~Ib{\'a}{\~n}ez},
  \bibinfo{author}{J.~Papavassiliou}, \bibinfo{journal}{Phys. Rev. D}
  \bibinfo{volume}{86} (\bibinfo{year}{2012}) \bibinfo{pages}{085033}.
\bibitem[{Strauss et~al.(2012)Strauss, Fischer, and
  Kellermann}]{Strauss:2012dg}
\bibinfo{author}{S.~Strauss}, \bibinfo{author}{C.~S. Fischer},
  \bibinfo{author}{C.~Kellermann}, \bibinfo{journal}{Phys. Rev. Lett.}
  \bibinfo{volume}{109} (\bibinfo{year}{2012}) \bibinfo{pages}{252001}.
\bibitem[{Meyers and Swanson(2014)}]{Meyers:2014iwa}
\bibinfo{author}{J.~Meyers}, \bibinfo{author}{E.~S. Swanson},
  \bibinfo{journal}{Phys. Rev. D} \bibinfo{volume}{90} (\bibinfo{year}{2014})
  \bibinfo{pages}{045037}.
\bibitem[{Maris and Roberts(2003)}]{Maris:2003vk}
\bibinfo{author}{P.~Maris}, \bibinfo{author}{C.~D. Roberts},
  \bibinfo{journal}{Int. J. Mod. Phys. E} \bibinfo{volume}{12}
  (\bibinfo{year}{2003}) \bibinfo{pages}{297--365}.
\bibitem[{Chang et~al.(2011)Chang, Roberts, and Tandy}]{Chang:2011vu}
\bibinfo{author}{L.~Chang}, \bibinfo{author}{C.~D. Roberts},
  \bibinfo{author}{P.~C. Tandy}, \bibinfo{journal}{Chin. J. Phys.}
  \bibinfo{volume}{49} (\bibinfo{year}{2011}) \bibinfo{pages}{955--1004}.
\bibitem[{Bashir et~al.(2012)}]{Bashir:2012fs}
\bibinfo{author}{A.~Bashir}, et~al., \bibinfo{journal}{Commun. Theor. Phys.}
  \bibinfo{volume}{58} (\bibinfo{year}{2012}) \bibinfo{pages}{79--134}.
\bibitem[{Clo{\"e}t and Roberts(2014)}]{Cloet:2013jya}
\bibinfo{author}{I.~C. Clo{\"e}t}, \bibinfo{author}{C.~D. Roberts},
  \bibinfo{journal}{Prog. Part. Nucl. Phys.} \bibinfo{volume}{77}
  (\bibinfo{year}{2014}) \bibinfo{pages}{1--69}.
\bibitem[{Binosi et~al.(2015)Binosi, Chang, Papavassiliou, and
  Roberts}]{Binosi:2014aea}
\bibinfo{author}{D.~Binosi}, \bibinfo{author}{L.~Chang},
  \bibinfo{author}{J.~Papavassiliou}, \bibinfo{author}{C.~D. Roberts},
  \bibinfo{journal}{Phys. Lett. B} \bibinfo{volume}{742} (\bibinfo{year}{2015})
  \bibinfo{pages}{183--188}.
\bibitem[{Wilson et~al.(2012)Wilson, Clo{\"e}t, Chang, and
  Roberts}]{Wilson:2011aa}
\bibinfo{author}{D.~J. Wilson}, \bibinfo{author}{I.~C. Clo{\"e}t},
  \bibinfo{author}{L.~Chang}, \bibinfo{author}{C.~D. Roberts},
  \bibinfo{journal}{Phys. Rev. C} \bibinfo{volume}{85} (\bibinfo{year}{2012})
  \bibinfo{pages}{025205}.
\bibitem[{Clo{\"e}t et~al.(2014)Clo{\"e}t, Bentz, and Thomas}]{Cloet:2014rja}
\bibinfo{author}{I.~C. Clo{\"e}t}, \bibinfo{author}{W.~Bentz},
  \bibinfo{author}{A.~W. Thomas}, \bibinfo{journal}{Phys. Rev. C}
  \bibinfo{volume}{90} (\bibinfo{year}{2014}) \bibinfo{pages}{045202}.
\bibitem[{Lichtenberg and Tassie(1967)}]{Lichtenberg:1967zz}
\bibinfo{author}{D.~B. Lichtenberg}, \bibinfo{author}{L.~J. Tassie},
  \bibinfo{journal}{Phys. Rev.} \bibinfo{volume}{155} (\bibinfo{year}{1967})
  \bibinfo{pages}{1601--1606}.
\bibitem[{Lichtenberg et~al.(1968)Lichtenberg, Tassie, and
  Keleman}]{Lichtenberg:1968zz}
\bibinfo{author}{D.~B. Lichtenberg}, \bibinfo{author}{L.~J. Tassie},
  \bibinfo{author}{P.~J. Keleman}, \bibinfo{journal}{Phys. Rev.}
  \bibinfo{volume}{167} (\bibinfo{year}{1968}) \bibinfo{pages}{1535--1542}.
\bibitem[{Ripani et~al.(2003)}]{Ripani:2002ss}
\bibinfo{author}{M.~Ripani}, et~al., \bibinfo{journal}{Phys. Rev. Lett.}
  \bibinfo{volume}{91} (\bibinfo{year}{2003}) \bibinfo{pages}{022002}.
\bibitem[{Burkert(2012)}]{Burkert:2012ee}
\bibinfo{author}{V.~D. Burkert}, \bibinfo{journal}{EPJ Web Conf.}
  \bibinfo{volume}{37} (\bibinfo{year}{2012}) \bibinfo{pages}{01017}.
\bibitem[{Kamano et~al.(2013)Kamano, Nakamura, Lee, and Sato}]{Kamano:2013iva}
\bibinfo{author}{H.~Kamano}, \bibinfo{author}{S.~X. Nakamura},
  \bibinfo{author}{T.~S.~H. Lee}, \bibinfo{author}{T.~Sato},
  \bibinfo{journal}{Phys. Rev. C} \bibinfo{volume}{88} (\bibinfo{year}{2013})
  \bibinfo{pages}{035209}.
\bibitem[{Chen et~al.(2012)}]{Chen:2012qrS}
\bibinfo{author}{C.~Chen}, et~al., \bibinfo{journal}{Few Body Syst.}
  \bibinfo{volume}{53} (\bibinfo{year}{2012}) \bibinfo{pages}{293--326}.
\bibitem[{Edwards et~al.(2011)Edwards, Dudek, Richards, and
  Wallace}]{Edwards:2011jj}
\bibinfo{author}{R.~G. Edwards}, \bibinfo{author}{J.~J. Dudek},
  \bibinfo{author}{D.~G. Richards}, \bibinfo{author}{S.~J. Wallace},
  \bibinfo{journal}{Phys. Rev. D} \bibinfo{volume}{84} (\bibinfo{year}{2011})
  \bibinfo{pages}{074508}.
\bibitem[{Chang et~al.(2011)Chang, Liu, and Roberts}]{Chang:2010hb}
\bibinfo{author}{L.~Chang}, \bibinfo{author}{Y.-X. Liu}, \bibinfo{author}{C.~D.
  Roberts}, \bibinfo{journal}{Phys. Rev. Lett.} \bibinfo{volume}{106}
  (\bibinfo{year}{2011}) \bibinfo{pages}{072001}.
\bibitem[{Riordan et~al.(2010)Riordan, Abrahamyan, Craver, Kelleher, Kolarkar
  et~al.}]{Riordan:2010id}
\bibinfo{author}{S.~Riordan}, \bibinfo{author}{S.~Abrahamyan},
  \bibinfo{author}{B.~Craver}, \bibinfo{author}{A.~Kelleher},
  \bibinfo{author}{A.~Kolarkar}, et~al., \bibinfo{journal}{Phys. Rev. Lett.}
  \bibinfo{volume}{105} (\bibinfo{year}{2010}) \bibinfo{pages}{262302}.
\bibitem[{Cates et~al.(2011)Cates, de~Jager, Riordan, and
  Wojtsekhowski}]{Cates:2011pz}
\bibinfo{author}{G.~Cates}, \bibinfo{author}{C.~de~Jager},
  \bibinfo{author}{S.~Riordan}, \bibinfo{author}{B.~Wojtsekhowski},
  \bibinfo{journal}{Phys. Rev. Lett.} \bibinfo{volume}{106}
  (\bibinfo{year}{2011}) \bibinfo{pages}{252003}.
\bibitem[{Qattan and Arrington(2012)}]{Qattan:2012zf}
\bibinfo{author}{I.~A. Qattan}, \bibinfo{author}{J.~Arrington},
  \bibinfo{journal}{Phys. Rev. C} \bibinfo{volume}{86} (\bibinfo{year}{2012})
  \bibinfo{pages}{065210}.
\bibitem[{Shanahan et~al.(2015)}]{Shanahan:2014tja}
\bibinfo{author}{P.~E. Shanahan}, et~al., \bibinfo{journal}{Phys. Rev. Lett.}
  \bibinfo{volume}{114} (\bibinfo{year}{2015}) \bibinfo{pages}{091802}.
\bibitem[{Aznauryan et~al.(2013)Aznauryan, Bashir, Braun, Brodsky, Burkert
  et~al.}]{Aznauryan:2012ba}
\bibinfo{author}{I.~Aznauryan}, \bibinfo{author}{A.~Bashir},
  \bibinfo{author}{V.~Braun}, \bibinfo{author}{S.~Brodsky},
  \bibinfo{author}{V.~Burkert}, et~al., \bibinfo{journal}{Int. J. Mod. Phys. E}
  \bibinfo{volume}{22} (\bibinfo{year}{2013}) \bibinfo{pages}{1330015}.
\bibitem[{Gothe et~al.(tion)}]{Gothe:2009}
\bibinfo{author}{R.~W. Gothe}, et~al., \bibinfo{year}{CLAS Collaboration}.
  \bibinfo{note}{{\emph{Nucleon Resonance Studies with CLAS12}, Jlab Experiment
  E12-09-003}}.
\bibitem[{Carman et~al.(tion)}]{Carman:2014}
\bibinfo{author}{D.~S. Carman}, et~al., \bibinfo{year}{CLAS Collaboration}.
  \bibinfo{note}{{\emph{Exclusive \mbox{$N^\ast \to K Y$} Studies with CLAS12},
  Jlab Experiment E12-06-108A}}.
\bibitem[{Jones et~al.(2000)}]{Jones:1999rz}
\bibinfo{author}{M.~K. Jones}, et~al., \bibinfo{journal}{Phys. Rev. Lett.}
  \bibinfo{volume}{84} (\bibinfo{year}{2000}) \bibinfo{pages}{1398--1402}.
\bibitem[{Gayou et~al.(2002)}]{Gayou:2001qd}
\bibinfo{author}{O.~Gayou}, et~al., \bibinfo{journal}{Phys. Rev. Lett.}
  \bibinfo{volume}{88} (\bibinfo{year}{2002}) \bibinfo{pages}{092301}.
\bibitem[{Gayou et~al.(2001)Gayou, Wijesooriya, Afanasev, Amarian, Aniol
  et~al.}]{Gayou:2001qt}
\bibinfo{author}{O.~Gayou}, \bibinfo{author}{K.~Wijesooriya},
  \bibinfo{author}{A.~Afanasev}, \bibinfo{author}{M.~Amarian},
  \bibinfo{author}{K.~Aniol}, et~al., \bibinfo{journal}{Phys. Rev. C}
  \bibinfo{volume}{64} (\bibinfo{year}{2001}) \bibinfo{pages}{038202}.
\bibitem[{Punjabi et~al.(2005)Punjabi, Perdrisat, Aniol, Baker, Berthot
  et~al.}]{Punjabi:2005wq}
\bibinfo{author}{V.~Punjabi}, \bibinfo{author}{C.~Perdrisat},
  \bibinfo{author}{K.~Aniol}, \bibinfo{author}{F.~Baker},
  \bibinfo{author}{J.~Berthot}, et~al., \bibinfo{journal}{Phys. Rev. C}
  \bibinfo{volume}{71} (\bibinfo{year}{2005}) \bibinfo{pages}{055202}.
\bibitem[{Puckett et~al.(2010)}]{Puckett:2010ac}
\bibinfo{author}{A.~J.~R. Puckett}, et~al., \bibinfo{journal}{Phys. Rev. Lett.}
  \bibinfo{volume}{104} (\bibinfo{year}{2010}) \bibinfo{pages}{242301}.
\bibitem[{Puckett et~al.(2012)Puckett, Brash, Gayou, Jones, Pentchev
  et~al.}]{Puckett:2011xg}
\bibinfo{author}{A.~J.~R. Puckett}, \bibinfo{author}{E.~J. Brash},
  \bibinfo{author}{O.~Gayou}, \bibinfo{author}{M.~K. Jones},
  \bibinfo{author}{L.~Pentchev}, et~al., \bibinfo{journal}{Phys. Rev. C}
  \bibinfo{volume}{85} (\bibinfo{year}{2012}) \bibinfo{pages}{045203}.
\bibitem[{Zhu et~al.(2001)}]{Zhu:2001md}
\bibinfo{author}{H.~Zhu}, et~al., \bibinfo{journal}{Phys. Rev. Lett.}
  \bibinfo{volume}{87} (\bibinfo{year}{2001}) \bibinfo{pages}{081801}.
\bibitem[{Bermuth et~al.(2003)Bermuth, Merle, Carasco, Baumann, Bohm
  et~al.}]{Bermuth:2003qh}
\bibinfo{author}{J.~Bermuth}, \bibinfo{author}{P.~Merle},
  \bibinfo{author}{C.~Carasco}, \bibinfo{author}{D.~Baumann},
  \bibinfo{author}{D.~Bohm}, et~al., \bibinfo{journal}{Phys. Lett. B}
  \bibinfo{volume}{564} (\bibinfo{year}{2003}) \bibinfo{pages}{199--204}.
\bibitem[{Warren et~al.(2004)}]{Warren:2003ma}
\bibinfo{author}{G.~Warren}, et~al., \bibinfo{journal}{Phys. Rev. Lett.}
  \bibinfo{volume}{92} (\bibinfo{year}{2004}) \bibinfo{pages}{042301}.
\bibitem[{Glazier et~al.(2005)}]{Glazier:2004ny}
\bibinfo{author}{D.~Glazier}, et~al., \bibinfo{journal}{Eur. Phys. J. A}
  \bibinfo{volume}{24} (\bibinfo{year}{2005}) \bibinfo{pages}{101--109}.
\bibitem[{Plaster et~al.(2006)}]{Plaster:2005cx}
\bibinfo{author}{B.~Plaster}, et~al., \bibinfo{journal}{Phys. Rev. C}
  \bibinfo{volume}{73} (\bibinfo{year}{2006}) \bibinfo{pages}{025205}.
\bibitem[{Clo{\"e}t et~al.(2013)Clo{\"e}t, Roberts, and Thomas}]{Cloet:2013gva}
\bibinfo{author}{I.~C. Clo{\"e}t}, \bibinfo{author}{C.~D. Roberts},
  \bibinfo{author}{A.~W. Thomas}, \bibinfo{journal}{Phys. Rev. Lett.}
  \bibinfo{volume}{111} (\bibinfo{year}{2013}) \bibinfo{pages}{101803}.
\bibitem[{Miller and Frank(2002)}]{Miller:2002qb}
\bibinfo{author}{G.~A. Miller}, \bibinfo{author}{M.~R. Frank},
  \bibinfo{journal}{Phys. Rev. C} \bibinfo{volume}{65} (\bibinfo{year}{2002})
  \bibinfo{pages}{065205}.
\bibitem[{Roberts et~al.(2013)Roberts, Holt, and Schmidt}]{Roberts:2013mja}
\bibinfo{author}{C.~D. Roberts}, \bibinfo{author}{R.~J. Holt},
  \bibinfo{author}{S.~M. Schmidt}, \bibinfo{journal}{Phys. Lett. B}
  \bibinfo{volume}{727} (\bibinfo{year}{2013}) \bibinfo{pages}{249--254}.
\bibitem[{Clo{\"e}t et~al.(l th)Clo{\"e}t, Krassnigg, and
  Roberts}]{Cloet:2007piS}
\bibinfo{author}{I.~C. Clo{\"e}t}, \bibinfo{author}{A.~Krassnigg},
  \bibinfo{author}{C.~D. Roberts}  (\bibinfo{year}{arXiv:0710.5746 [nucl-th]}).
  \bibinfo{note}{{\emph{Dynamics, Symmetries and Hadron Properties}}}.
\bibitem[{Roberts et~al.(l th)Roberts, Chang, Clo{\"e}t, and
  Roberts}]{Roberts:2010huS}
\bibinfo{author}{H.~L.~L. Roberts}, \bibinfo{author}{L.~Chang},
  \bibinfo{author}{I.~C. Clo{\"e}t}, \bibinfo{author}{C.~D. Roberts}
  (\bibinfo{year}{arXiv:1007.3566 [nucl-th]}). \bibinfo{note}{{\emph{Exposing
  the dressed quark's mass}}}.
\bibitem[{Clo{\"e}t et~al.(2011)Clo{\"e}t, Roberts, and Wilson}]{Cloet:2011qu}
\bibinfo{author}{I.~C. Clo{\"e}t}, \bibinfo{author}{C.~D. Roberts},
  \bibinfo{author}{D.~J. Wilson}, \bibinfo{journal}{AIP Conf. Proc.}
  \bibinfo{volume}{1388} (\bibinfo{year}{2011}) \bibinfo{pages}{121--127}.
\bibitem[{Parno et~al.(2015)}]{Parno:2014xzb}
\bibinfo{author}{D.~Parno}, et~al., \bibinfo{journal}{Phys. Lett. B}
  \bibinfo{volume}{744} (\bibinfo{year}{2015}) \bibinfo{pages}{309--314}.
\bibitem[{Clo{\"e}t and Miller(2012)}]{Cloet:2012cy}
\bibinfo{author}{I.~C. Clo{\"e}t}, \bibinfo{author}{G.~A. Miller},
  \bibinfo{journal}{Phys. Rev. C} \bibinfo{volume}{86} (\bibinfo{year}{2012})
  \bibinfo{pages}{015208}.

\end{thebibliography}

\end{document}